\documentclass[aps,prl,preprint,tightenlines,superscriptaddress,showpacs,byrevtex,amsmath,amssymb]{revtex4-1}

\usepackage{graphicx} 
\usepackage{dcolumn}  
\usepackage[colorlinks]{hyperref}
\usepackage[hyphenbreaks]{breakurl}

\graphicspath{{ps}}

\renewcommand{\arraystretch}{1.1}

\begin{document}




\title{ \quad\\[0.5cm] Measurements of isospin asymmetry and difference of direct $CP$ asymmetries \\in inclusive {\boldmath $B \to X_s \gamma$} decays}


\noaffiliation
\affiliation{University of the Basque Country UPV/EHU, 48080 Bilbao}
\affiliation{Beihang University, Beijing 100191}
\affiliation{University of Bonn, 53115 Bonn}
\affiliation{Brookhaven National Laboratory, Upton, New York 11973}
\affiliation{Budker Institute of Nuclear Physics SB RAS, Novosibirsk 630090}
\affiliation{Faculty of Mathematics and Physics, Charles University, 121 16 Prague}
\affiliation{Chonnam National University, Kwangju 660-701}
\affiliation{University of Cincinnati, Cincinnati, Ohio 45221}
\affiliation{Deutsches Elektronen--Synchrotron, 22607 Hamburg}
\affiliation{Duke University, Durham, North Carolina 27708}
\affiliation{University of Florida, Gainesville, Florida 32611}
\affiliation{Key Laboratory of Nuclear Physics and Ion-beam Application (MOE) and Institute of Modern Physics, Fudan University, Shanghai 200443}
\affiliation{II. Physikalisches Institut, Georg-August-Universit\"at G\"ottingen, 37073 G\"ottingen}
\affiliation{SOKENDAI (The Graduate University for Advanced Studies), Hayama 240-0193}
\affiliation{Hanyang University, Seoul 133-791}
\affiliation{University of Hawaii, Honolulu, Hawaii 96822}
\affiliation{High Energy Accelerator Research Organization (KEK), Tsukuba 305-0801}
\affiliation{J-PARC Branch, KEK Theory Center, High Energy Accelerator Research Organization (KEK), Tsukuba 305-0801}
\affiliation{Forschungszentrum J\"{u}lich, 52425 J\"{u}lich}
\affiliation{IKERBASQUE, Basque Foundation for Science, 48013 Bilbao}
\affiliation{Indian Institute of Science Education and Research Mohali, SAS Nagar, 140306}
\affiliation{Indian Institute of Technology Guwahati, Assam 781039}
\affiliation{Indian Institute of Technology Hyderabad, Telangana 502285}
\affiliation{Indian Institute of Technology Madras, Chennai 600036}
\affiliation{Indiana University, Bloomington, Indiana 47408}
\affiliation{Institute of High Energy Physics, Chinese Academy of Sciences, Beijing 100049}
\affiliation{Institute of High Energy Physics, Vienna 1050}
\affiliation{Institute for High Energy Physics, Protvino 142281}
\affiliation{INFN - Sezione di Napoli, 80126 Napoli}
\affiliation{INFN - Sezione di Torino, 10125 Torino}
\affiliation{Advanced Science Research Center, Japan Atomic Energy Agency, Naka 319-1195}
\affiliation{J. Stefan Institute, 1000 Ljubljana}
\affiliation{Kanagawa University, Yokohama 221-8686}
\affiliation{Institut f\"ur Experimentelle Teilchenphysik, Karlsruher Institut f\"ur Technologie, 76131 Karlsruhe}
\affiliation{King Abdulaziz City for Science and Technology, Riyadh 11442}
\affiliation{Department of Physics, Faculty of Science, King Abdulaziz University, Jeddah 21589}
\affiliation{Kitasato University, Tokyo 108-0072}
\affiliation{Korea Institute of Science and Technology Information, Daejeon 305-806}
\affiliation{Korea University, Seoul 136-713}
\affiliation{Kyoto University, Kyoto 606-8502}
\affiliation{Kyungpook National University, Daegu 702-701}
\affiliation{LAL, Univ. Paris-Sud, CNRS/IN2P3, Universit\'{e} Paris-Saclay, Orsay}
\affiliation{\'Ecole Polytechnique F\'ed\'erale de Lausanne (EPFL), Lausanne 1015}
\affiliation{P.N. Lebedev Physical Institute of the Russian Academy of Sciences, Moscow 119991}
\affiliation{Faculty of Mathematics and Physics, University of Ljubljana, 1000 Ljubljana}
\affiliation{Ludwig Maximilians University, 80539 Munich}
\affiliation{Luther College, Decorah, Iowa 52101}
\affiliation{University of Maribor, 2000 Maribor}
\affiliation{Max-Planck-Institut f\"ur Physik, 80805 M\"unchen}
\affiliation{School of Physics, University of Melbourne, Victoria 3010}
\affiliation{University of Mississippi, University, Mississippi 38677}
\affiliation{University of Miyazaki, Miyazaki 889-2192}
\affiliation{Moscow Physical Engineering Institute, Moscow 115409}
\affiliation{Moscow Institute of Physics and Technology, Moscow Region 141700}
\affiliation{Graduate School of Science, Nagoya University, Nagoya 464-8602}
\affiliation{Kobayashi-Maskawa Institute, Nagoya University, Nagoya 464-8602}
\affiliation{Universit\`{a} di Napoli Federico II, 80055 Napoli}
\affiliation{Nara Women's University, Nara 630-8506}
\affiliation{National United University, Miao Li 36003}
\affiliation{Department of Physics, National Taiwan University, Taipei 10617}
\affiliation{H. Niewodniczanski Institute of Nuclear Physics, Krakow 31-342}
\affiliation{Nippon Dental University, Niigata 951-8580}
\affiliation{Niigata University, Niigata 950-2181}
\affiliation{University of Nova Gorica, 5000 Nova Gorica}
\affiliation{Novosibirsk State University, Novosibirsk 630090}
\affiliation{Osaka City University, Osaka 558-8585}
\affiliation{Pacific Northwest National Laboratory, Richland, Washington 99352}
\affiliation{Peking University, Beijing 100871}
\affiliation{University of Pittsburgh, Pittsburgh, Pennsylvania 15260}
\affiliation{Punjab Agricultural University, Ludhiana 141004}
\affiliation{Theoretical Research Division, Nishina Center, RIKEN, Saitama 351-0198}
\affiliation{University of Science and Technology of China, Hefei 230026}
\affiliation{Showa Pharmaceutical University, Tokyo 194-8543}
\affiliation{Soongsil University, Seoul 156-743}
\affiliation{Stefan Meyer Institute for Subatomic Physics, Vienna 1090}
\affiliation{Sungkyunkwan University, Suwon 440-746}
\affiliation{School of Physics, University of Sydney, New South Wales 2006}
\affiliation{Department of Physics, Faculty of Science, University of Tabuk, Tabuk 71451}
\affiliation{Tata Institute of Fundamental Research, Mumbai 400005}
\affiliation{Department of Physics, Technische Universit\"at M\"unchen, 85748 Garching}
\affiliation{Toho University, Funabashi 274-8510}
\affiliation{Department of Physics, Tohoku University, Sendai 980-8578}
\affiliation{Earthquake Research Institute, University of Tokyo, Tokyo 113-0032}
\affiliation{Department of Physics, University of Tokyo, Tokyo 113-0033}
\affiliation{Tokyo Institute of Technology, Tokyo 152-8550}
\affiliation{Tokyo Metropolitan University, Tokyo 192-0397}
\affiliation{Virginia Polytechnic Institute and State University, Blacksburg, Virginia 24061}
\affiliation{Wayne State University, Detroit, Michigan 48202}
\affiliation{Yamagata University, Yamagata 990-8560}
\affiliation{Yonsei University, Seoul 120-749}
  \author{S.~Watanuki}\affiliation{Department of Physics, Tohoku University, Sendai 980-8578} 
  \author{A.~Ishikawa}\affiliation{Department of Physics, Tohoku University, Sendai 980-8578} 
  \author{I.~Adachi}\affiliation{High Energy Accelerator Research Organization (KEK), Tsukuba 305-0801}\affiliation{SOKENDAI (The Graduate University for Advanced Studies), Hayama 240-0193} 
  \author{H.~Aihara}\affiliation{Department of Physics, University of Tokyo, Tokyo 113-0033} 
  \author{S.~Al~Said}\affiliation{Department of Physics, Faculty of Science, University of Tabuk, Tabuk 71451}\affiliation{Department of Physics, Faculty of Science, King Abdulaziz University, Jeddah 21589} 
  \author{D.~M.~Asner}\affiliation{Brookhaven National Laboratory, Upton, New York 11973} 
  \author{T.~Aushev}\affiliation{Moscow Institute of Physics and Technology, Moscow Region 141700} 
  \author{R.~Ayad}\affiliation{Department of Physics, Faculty of Science, University of Tabuk, Tabuk 71451} 
  \author{V.~Babu}\affiliation{Tata Institute of Fundamental Research, Mumbai 400005} 
  \author{I.~Badhrees}\affiliation{Department of Physics, Faculty of Science, University of Tabuk, Tabuk 71451}\affiliation{King Abdulaziz City for Science and Technology, Riyadh 11442} 
% ...
\collaboration{The Belle Collaboration}


\begin{abstract}
We report measurements of isospin asymmetry $\Delta_{0-}$ and difference of direct $CP$ asymmetries $\Delta A_{CP}$ between charged and neutral $B \to X_s \gamma$ decays. 
This analysis is based on the data sample containing $772 \times 10^6 B\bar{B}$ pairs 
that was collected with the Belle detector at the KEKB energy-asymmetric $e^+ e^-$ collider.
Using a sum-of-exclusive technique with invariant $X_s$ mass up to 2.8~GeV/$c^2$, we obtain
$\Delta_{0-} = \bigl[-0.48 \pm 1.49 {\rm (stat.)} \pm 0.97 {\rm (syst.)} \pm 1.15 {(f_{+-}/f_{00})}\bigr]$\% and
$\Delta A_{CP} = \bigl[+3.69 \pm 2.65 {\rm (stat.)} \pm 0.76{\rm (syst.)}\bigr]$\%,
where the last uncertainty for $\Delta_{0-}$ is due to the uncertainty on the production ratio of $B^+B^-$ to $B^0\bar{B}^0$ in $\Upsilon(4S)$ decays.
The measured value of $\Delta_{0-}$ is consistent with zero, allowing us to constrain the resolved photon contribution in the $B \to X_s \gamma$, and improve the branching fraction prediction. The result for $\Delta A_{CP}$ is consistent with the prediction of the SM. We also measure the direct $CP$ asymmetries for charged and neutral $B \to X_s \gamma$ decays.
All the measurements are the most precise to date.
\end{abstract}

\maketitle

\tighten

{\renewcommand{\thefootnote}{\fnsymbol{footnote}}}
\setcounter{footnote}{0}


\section{I. Introduction}
The radiative $b \to s \gamma$ decay proceeds predominantly via one-loop electromagnetic penguin diagrams at the lowest order in the standard model~(SM). This decay is sensitive to new physics~(NP), which can alter the branching fraction, or direct $CP$ asymmetry defined as 
\begin{eqnarray}
A_{CP}        = \frac{\Gamma(\bar{B} \to \bar{X_s} \gamma)-\Gamma(B \to X_s \gamma)}{\Gamma(\bar{B} \to \bar{X_s} \gamma)+\Gamma(B \to X_s \gamma)},
\end{eqnarray}
where $\Gamma$ denotes the partial width.

Precision measurements of $B \to X_s \gamma$ branching fraction~${\cal{B}}(B \to X_s \gamma)$
~\cite{Chen:2001fja,Aubert:2007my,Lees:2012ym,Lees:2012wg,Limosani:2009qg,Saito:2014das}
are in good agreement with the SM prediction~\cite{Misiak:2015xwa} and set a strong constraint on NP models~\cite{BFNP}. The theoretical uncertainty in the prediction of ${\cal{B}}(B \to X_s \gamma)$ is about 7\% which is comparable with the experimental uncertainty of the current world average~\cite{PDG2018}. The Belle~II experiment is expected to measure the branching fraction with a precision of about 3\%~\cite{Kou:2018nap}. Thus, the reduction of the theoretical uncertainty is crucial to further constrain NP models. The largest uncertainty in the theoretical prediction is due to non-perturbative effects, one of which is the resolved photon contributions~\cite{Lee:2006wn}. Since the resolved photon contribution from a hard gluon and a light quark scattering to the $B \to X_s \gamma$ branching fraction~(${\cal{B}}_{\rm RP}^{78}$) depends on the charge of the light quark and can be hence related to the isospin asymmetry in $B \to X_s \gamma$~($\Delta_{0-}$) as~\cite{Lee:2006wn,Misiak:2009nr,Benzke:2010js}
\begin{eqnarray}
\frac{{\cal{B}}_{\rm RP}^{78}}{\cal{B}} &\simeq& -\frac{(1\pm0.3)}{3}\Delta_{0-},
\label{eqn:rp}
\end{eqnarray}
where the uncertainty of $\pm0.3$ in the right-hand side is associated with $SU(3)$ flavor-symmetry breaking. The isospin asymmetry is defined as 
\begin{eqnarray}
\nonumber
\Delta_{0-}   &=& \frac{\Gamma(\bar{B}^0 \to X_s^0 \gamma)-\Gamma(B^- \to X_s^- \gamma)}{\Gamma(\bar{B}^0 \to X_s^0 \gamma)+\Gamma(B^- \to X_s^- \gamma)}\\ 
&=& \frac{\frac{\tau_{B^-}}{\tau_{\bar{B}^0}}\frac{f_{+-}}{f_{00}} N(\bar{B}^0 \to X_s^0 \gamma) - N(B^- \to X_s^- \gamma)}{\frac{\tau_{B^-}}{\tau_{\bar{B}^0}}\frac{f_{+-}}{f_{00}} N(\bar{B}^0 \to X_s^0 \gamma) + N(B^- \to X_s^- \gamma)},
\label{eqn:iso}
\end{eqnarray}
where $N$ is the number of produced signal events including charge-conjugate decays, $\tau_{B^-}/\tau_{\bar{B}^0}=\tau_{B^+}/\tau_{B^0}$ is the lifetime ratio of $B^+$ to $B^0$ mesons, $f_{+-}$ and $f_{00}$ are the production ratio of $B^+B^-$ to $B^0\bar{B}^0$ in $\Upsilon(4S)$ decays, respectively. If the measured value of $\Delta_{0-}$ is consistent with zero, the resolved photon contribution is small and reducing in the theoretical uncertainty on ${\cal{B}}(B \to X_s \gamma)$. Recently, evidence for isospin violation in exclusive $B \to K^{\ast}(892) \gamma$~($\Delta_{0+}$) has been reported~\cite{Horiguchi:2017ntw} where the measured value, $\Delta_{0+} = (+6.2\pm1.5\pm0.5\pm1.2)\%$, is consistent with SM predictions~\cite{Kagan:2001zk,Keum:2004is,Ball:2006eu,Jung:2012vu,Ahmady:2013cva,Lyon:2013gba}. If the isospin asymmetry for the inclusive decays is consistent with this value, the resolved photon contribution to $B \to X_s \gamma$ decays could be sizable.

The direct $CP$ asymmetry in $B \to X_s \gamma$ is also a sensitive probe for NP~\cite{Wolfenstein:1994jw,Asatrian:1996as,Kagan:1998bh,Chua:1998dx,Kiers:2000xy,Asatryan:2000kt,Hurth:2003dk,Jung:2012vu,Barbieri:2011fc,Shimizu:2012ru,Arbey:2014msa}. Belle~\cite{Nishida:2003paa} and BaBar~\cite{Lees:2014uoa} measured this quantity, and the current world average $(+1.5\pm2.0)$\%~\cite{PDG2018} is in agreement and of comparable precision, with the SM prediction, $-0.6\% < A_{CP}^{\rm SM} < +2.8\%$~\cite{Benzke:2010tq}. The dominant theoretical uncertainty is due to the limited knowledge of the resolved photon contributions. A newly proposed observable is the difference of the direct $CP$ asymmetries between the charged and neutral $B$ mesons,~$\Delta A_{CP} = A_{CP}(B^+ \to X_s^+ \gamma) - A_{CP}(B^0 \to X_s^0 \gamma)$, where terms with large weak phase in the SM cancel out, and only the spectator-quark-flavor dependent term representing interference between electromagnetic and chromomagnetic dipole operators survives~\cite{Benzke:2010tq}:
\begin{eqnarray}
\nonumber
\Delta A_{CP} 
 &=& 4\pi^2\alpha_s \frac{\tilde \Lambda_{78}}{m_b} {\rm Im}\bigg(\frac{C_8}{C_7}\bigg)\\
&\approx& 0.12\bigg(\frac{\tilde \Lambda_{78}}{100\rm~MeV}\bigg) {\rm Im}\bigg(\frac{C_8}{C_7}\bigg),
\label{eqn:dacp}
\end{eqnarray}
where $\alpha_s$ is the strong coupling constant, $\tilde \Lambda_{78}$ is the hadronic parameter denoting the interference between electromagnetic and chromomagnetic dipole diagrams, $m_b$ is the bottom quark mass, and $C_7$ and $C_8$ are the Wilson coefficients for electromagnetic and chromomagnetic dipole operators, respectively~\cite{Czakon:2015exa}.
In the SM, $C_7$ and $C_8$ are both real, therefore $\Delta A_{CP}$ is zero, but in several NP models $\Delta A_{CP}$ can reach the level of 10\% in magnitude~\cite{Benzke:2010tq,Malm:2015oda,Endo:2017ums}.


BaBar measured $\Delta_{0-}$ and $\Delta A_{CP}$ using data samples of 81.9~fb$^{-1}$ and 429~fb$^{-1}$, respectively, as $\Delta_{0-} = (-0.6\pm5.8\pm0.9\pm2.4)$\%~\cite{Aubert:2005cua} and $\Delta A_{CP}=(+5.0\pm3.9\pm1.5)$\%~\cite{Lees:2014uoa}, where the first uncertainty is statistical, the second is systematic, and the last one for $\Delta_{0-}$ is due to the uncertainty on the fraction of $B^+B^-$ to $B^0\bar{B}^0$ production in $\Upsilon(4S)$ decays. The precisions are limited by statistical uncertainties. Improving these measurements is highly desirable to reduce the theoretical uncertainty of ${\cal{B}}(B \to X_s \gamma)$ in the SM as well as to search for NP.

In this article, we report first measurements of $\Delta_{0-}$ and $\Delta A_{CP}$ in inclusive $B \to X_s \gamma$ at Belle assuming that the two observables have no dependence on decay modes nor on the invariant mass of the $X_s$ system~($M_{X_s}$).
In addition, we present measurements of individual $A_{CP}$ for the charged and neutral decay and their average with $\bar{A}_{CP} = (A_{CP}(B^- \to X_s \gamma) + A_{CP}(\bar{B}^0 \to X_s \gamma))/2$.
All measurements are based on the full data sample of 711~fb$^{-1}$, containing ${772\times 10^6 B\bar{B}}$ pairs, recorded at the $\Upsilon(4S)$ resonance~(on-resonance data) with the Belle detector~\cite{Belle} at the KEKB $e^+ e^-$ collider~\cite{KEKB}. In addition, the data sample of 89~fb$^{-1}$ accumulated 60~MeV below the $\Upsilon(4S)$ peak~(off-resonance data), which is below the $B\bar{B}$ production threshold, is used to provide a background description.
The result for $A_{CP}(B \to X_s \gamma)$ supersedes our previous measurement~\cite{Nishida:2003paa}.

\section{II. Belle Detector}

The Belle detector is a large-solid-angle magnetic
spectrometer that consists of a silicon vertex detector (SVD),
a 50-layer central drift chamber (CDC), an array of
aerogel threshold Cherenkov counters (ACC),  
a barrel-like arrangement of time-of-flight
scintillation counters (TOF), and an electromagnetic calorimeter
comprised of CsI(Tl) crystals (ECL). All the sub-detectors are located inside 
a superconducting solenoid coil that provides a 1.5~T
magnetic field.  An iron flux-return placed outside of
the coil is instrumented to detect $K_L^0$ mesons and muons. 
The $z$ axis is aligned with the direction opposite the $e^+$ beam.
The detector is described in detail elsewhere~\cite{Belle}.

\section{III. MC Simulation}
The selection is optimized with Monte Carlo~(MC) simulation samples. The MC simulation events are generated with EvtGen~\cite{EvtGen} and the detector simulation is done with GEANT3~\cite{GEANT}. 
We generate two types of signal MC simulation samples, according to the $X_s$ mass region: in the region $M_{X_s} <$\,1.15\,GeV/$c^2$, the $X_s$ system solely consists of $K^{\ast}(892)$ while in the region $M_{X_s} >$ 1.15\,GeV/$c^2$, $X_s$ system is simulated inclusively without any specific resonances, except for $K_2^{\ast}(1430)$.

In the inclusive signal MC simulation sample, various resonances and final states are simulated.
The photon energy spectrum in this sample is produced following the Kagan-Neubert model~\cite{Kagan:1998ym}. 
The model has two parameters: the $b$ quark mass ($m_b$) and the Fermi-motion parameter of the $b$ quark inside the $B$ meson ($\mu_{\pi}^2$).
The nominal values of these parameters are determined from a fit to the Belle inclusive photon energy spectrum \cite{Limosani:2009qg}: $m_b$\,=\,4.440 GeV/$c^2$ and  $\mu_{\pi}^2$\,=\,0.750 GeV$^2$.
Further, the generated light quark pair is fragmented into final-state hadrons using PYTHIA \cite{Pythia}.

Since the $B \to K_2^{\ast}(1430)\gamma$ decay has a relatively large branching fraction, dedicated MC simulation samples are generated. The decay is generated with the measured branching fraction
and then added to the inclusive signal MC simulation sample. To match the photon spectrum with the theoretical one, the $M_{X_s}$ distribution for $K\pi$ and $K2\pi$ modes in the inclusive sample is rescaled.
The signal reconstruction efficiency depends on the particle content in the final state; thus, the hadronization of $X_s$ is studied using data.
We set the branching fraction of $B \to X_s \gamma$ to the current world average~\cite{PDG2018} in order to optimize the background rejection.

\section{IV. Event Selection}
We reconstruct $B \to X_s \gamma$ decays with 38 exclusive $X_s$ final states listed in Table~\ref{tab:recmode}.
As shown in Table~\ref{tab:modecategory}, we group the final states into ten categories for the purpose of specific selections and fragmentation model calibrations.
The reconstructed decay modes cover 59\% of the total $X_s$ rate, according to the MC simulations. Assuming the $K^0$ meson to decay equally into $K^0_L$ and $K^0_S$, the proportion of our measured final states is 77\% of the total $X_s$ rate. For neutral $B$ decays, all flavor-specific final states are used for the measurements of both $A_{CP}$ and $\Delta_{0-}$, and 11 flavor-non-specific final states, denoted as $B_{\rm fns}$, are only used for the measurement of $\Delta_{0-}$.


\begin{table}[b]
	\caption{ Reconstructed $X_s$ final states~\cite{CC}. The mode IDs with an asterisk indicate the flavor-non-specific decays which are not used for $A_{CP}$ measurements.}
	\label{tab:recmode}
	\begin{tabular}
		{l|l|l|l}
		\hline \hline
		Mode ID & Final state & Mode ID & Final state \\
		\hline
		1   & $K^+\pi^- $                 & 20   & $K_S^0\pi^+\pi^0\pi^0$      \\
		2   & $K_S^0\pi^+$                & 21   & $K^{+}\pi^+\pi^-\pi^0\pi^0$   \\
		3   & $K^+\pi^0$                  & 22*  & $K_S^0\pi^+\pi^-\pi^0\pi^0$ \\
		4*  & $K_S^0\pi^0$                & 23   & $K^+\eta$      \\
		5   & $K^+\pi^+\pi^-$             & 24*  & $K_S^0\eta$      \\
		6*  & $K_S^0\pi^+\pi^-$           & 25   & $K^+\eta\pi^-$      \\
		7   & $K^+\pi^-\pi^0$             & 26   & $K_S^0\eta\pi^+$      \\
		8   & $K_S^0\pi^+\pi^0$           & 27   & $K^+\eta\pi^0$      \\
		9   & $K^{+}\pi^+\pi^-\pi^-$      & 28*  & $K_S^0\eta\pi^0$      \\
		10  & $K_S^0\pi^+\pi^+\pi^-$      & 29   & $K^+\eta\pi^+\pi^-$      \\
		11  & $K^+\pi^+\pi^-\pi^0$        & 30*  & $K_S^0\eta\pi^+\pi^-$      \\
		12* & $K_S^0\pi^+\pi^-\pi^0$      & 31   & $K^+\eta\pi^-\pi^0$      \\
		13  & $K^{+}\pi^+\pi^+\pi^-\pi^-$ & 32   & $K_S^0\eta\pi^+\pi^0$      \\
		14* & $K_S^0\pi^+\pi^+\pi^-\pi^-$ & 33   & $K^+K^+K^-$      \\
		15  & $K^+\pi^+\pi^-\pi^-\pi^0$   & 34*  & $K^+K^-K_S^0$      \\
		16  & $K_S^0\pi^+\pi^+\pi^-\pi^0$ & 35   & $K^+K^+K^-\pi^-$      \\
		17  & $K^{+}\pi^0\pi^0$           & 36   & $K^+K^-K_S^0\pi^+$      \\
		18* & $K_S^0\pi^0\pi^0$           & 37   & $K^+K^+K^-\pi^0$      \\
		19  & $K^+\pi^-\pi^0\pi^0$        & 38*  & $K^+K^-K_S^0\pi^0$      \\
		\hline \hline
	\end{tabular}
\end{table}

\begin{table}[tb]
	\begin{center}
		  \caption{Mode category definitions for $X_s$ fragmentation study.}
		  \label{tab:modecategory}
          \begin{tabular}{ccc} \hline \hline
              Mode category &  Definition   & Mode ID    \\ \hline 
              1    & $K\pi$ without $\pi^0$ &  1,2   \\ 
              2    & $K\pi$ with $\pi^0$    &  3,4   \\ 
              3    & $K2\pi$ without $\pi^0$&  5,6   \\ 
              4    & $K2\pi$ with $\pi^0$   &  7,8   \\ 
              5    & $K3\pi$ without $\pi^0$&  9,10   \\ 
              6    & $K3\pi$ with $\pi^0$   &  11,12   \\ 
              7    & $K4\pi$                &  13--16   \\ 
              8    & $K2\pi^0$               &  17--22   \\ 
              9    & $K\eta$                &  23--32   \\ 
              10   & 3$K$                   &  33--38   \\ \hline \hline
          \end{tabular}
	\end{center}
\end{table}

High-energy prompt photons are selected as isolated clusters
in the ECL that are not matched to any charged tracks reconstructed by the SVD and the CDC. 
The cluster energy in the center of mass~(CM) system is required to be between 1.5 and 3.4~GeV.
The polar angle of the photon direction must be within the barrel ECL, $33^{\circ}<\theta <\,132^{\circ}$.
We also require the cluster shape to be consistent with an electromagnetic shower, $E_{9}/E_{25}$ $>$ 0.95, where $E_{9}/E_{25}$ is the ratio of energy 
deposits in the 3 $\times$ 3 array of CsI(Tl) crystals 
to that in the 5 $\times$ 5 array centered on the crystal with maximum energy.
In order to reduce contaminations from asymmetric 
$\eta \to \gamma \gamma$ or $\pi^{0} \to \gamma \gamma$ decays, 
the photon candidate is paired with all other photons
in the event with energy greater than 40~MeV. We reject
the pairs based on likelihoods (${\cal{ L }}_{\pi^0}$ and ${\cal{ L }}_{\eta}$), constructed from their invariant mass, 
and the energy and polar angle of the additional photon in the CM system~\cite{Koppenburg:2004fz}.
The photon candidate which has ${\cal{ L }}_{\pi^0} >0.05$ or ${\cal{ L }}_{\eta}>0.10$ is discarded.

Charged particles, except pions from $K_S^0$ decays, are required to have a distance 
of closest approach to the interaction point~(IP) within $\pm 5.0$~cm along the $z$ axis and $\pm 0.5$~cm 
in the transverse $x$-$y$ plane, and  a laboratory momentum above 100~MeV/$c$.
Charged kaons and pions are identified based on a likelihood ratio 
constructed from the specific ionization measurements in the CDC, time-of-flight information from the TOF, 
and response from the ACC~\cite{Nakano:2002jw}.

Neutral kaon ($K_S^0$) candidates are reconstructed from pairs of oppositely-charged tracks, treated as pions, and identified by a
multivariate analysis~\cite{NB} based on two sets of input variables~\cite{nakano}. 
The first set that separates $K_S^0$ candidates from the combinatorial background are:
(1) the $K_S^0$ momentum in the laboratory frame,
(2) the distance along the $z$ axis between the two track helices at their closest approach,
(3) the flight length in the $x$-$y$ plane,
(4) the angle between the $K_S^0$ momentum and the vector joining its decay vertex to the nominal IP,
(5) the angle between the $\pi$ momentum and the laboratory-frame direction of the $K_S^0$ in its rest frame,
(6) the distances of closest approach in the $x$-$y$ plane between the IP and the pion helices,
(7) the numbers of hits for axial and stereo wires in the CDC for each pion,
and (8) the presence or absence of associated hits in the SVD for each pion.
The second set of variables, which identifies ${\Lambda \to p\pi^-}$ background that has a similar long-lived topology, are:
(1) particle identification information, momentum, and polar angles of the two daughter tracks in the laboratory frame,
and (2) the invariant mass calculated with the proton- and pion-mass hypotheses for the two tracks.
In total, the first and second sets comprise 13 and 7 input variables, respectively. 
The selected $K^0_S$ candidates are required to have an invariant mass within $\pm10$~MeV/$c^2$ of
the nominal value~\cite{PDG2018}, corresponding to a $\pm$3$\sigma$ interval in mass resolution, where $\sigma$ represents the standard deviation.

We reconstruct $\pi^0$ candidates from two photons each with energy greater than 50~MeV.
We require a minimum momentum of 100~MeV/$c$ in the CM frame and the invariant mass to be within $\pm10$~MeV/$c^2$ of the nominal $\pi^0$ mass, corresponding to about 1.5$\sigma$ in resolution. To reduce the large combinatorial background, we require the cosine of the angle between two photons in the CM frame to be greater than 0.5.

The $\eta$ candidates are formed from two photons, each with energy greater than 100~MeV.
The photon pairs with invariant mass satisfying 515~MeV$/c^2$ $<$ $M_{\gamma\gamma}$ $<$ 570~MeV$/c^2$, which corresponds to about 2$\sigma$ in resolution, are retained. We require a CM momentum to be
greater than 500~MeV/$c$ and an absolute value of the cosine of the helicity angle, which is the angle between momentum of one of the photons and direction of laboratory system in the $\eta$ rest frame, to be less than 0.8.

The 38 $X_s$ final states are reconstructed from the selected $\pi^+$, $\pi^0$, $K^+$, $K_S^0$, and $\eta$ candidates. In order to reduce the large combinatorial background from events with high multiplicity, we require $M_{X_s} < 2.8$~GeV$/c^2$, which corresponds to photon energy threshold of about $1.9$~GeV. The $K4\pi$ and $K2\pi^0$ mode categories, listed in Table~\ref{tab:modecategory}, have substantial background. 
Therefore, the momentum of the first and second leading pions (neutral pions) in $K4\pi$ ($K2\pi^0$) category is required to be above 400~MeV/$c$ and 250~MeV/$c$, respectively.

$B$ meson candidates are reconstructed by combining an $X_s$ with a prompt photon candidate. 
We form two kinematic variables to select $B$ mesons:
the energy difference ${\Delta E \equiv E_{B}^{\rm CM} - E_{\rm beam}^{\rm CM}}$
and the beam-energy constrained mass ${M_{\rm bc} \equiv  \sqrt{ (E_{\rm beam}^{\rm CM}/c^2)^2 - ({\bold{p}}_{B}^{\rm CM}/c)^2 }}$,
where $E_{\rm beam}^{\rm CM}$, $E_{B}^{\rm CM}$ and ${\bold p}_{B}^{\rm CM}$ are the beam energy, energy and momentum of the $B$ candidate in the CM system, respectively.
The $B$ momentum vector ${\bold{p}}_{B}^{\rm CM}$ is calculated without using the magnitude of the photon momentum according to ${{\bold{p}}_{B}^{\rm CM} = {\bold{p}}_{X_{s}}^{\rm CM} + {\bold{p}}_{\gamma}^{\rm CM}/|{\bold{p}}_{\gamma}^{\rm CM}| \times (E_{\rm beam}^{\rm CM} - E_{X_{s}}^{\rm CM})}$, as the $X_s$ momentum~(${\bold{p}}_{X_{s}}^{\rm CM}$) and the beam energy are determined with a substantially better precision than that of the photon candidate.
We define the signal region in $\Delta E$ and $M_{\rm bc}$ as $-0.15$~GeV $ < \Delta E <$ 0.08~GeV and 5.27~GeV$/c^2$ $< M_{\rm bc} <$ 5.29~GeV$/c^2$. 
The $\Delta E$ selection is tightened to $-$0.10\,GeV$<\Delta E<$0.05\,GeV for the final states with $2\pi^0$ and $\eta\pi^0$ (mode IDs 17--22, 27, 28, 31 and 32) due to larger combinatorial backgrounds.
To determine the signal yield and extract physics observables, we fit to the $M_{\rm bc}$ distribution in the wider range of 5.20~GeV$/c^2$ $< M_{\rm bc} <$ 5.29~GeV$/c^2$.

\section{V. Background Rejection}

After reconstructing the $B$ meson candidates, two dominant backgrounds still remain: events with $D$ meson decays and continuum $e^+e^- \to q\bar{q}$ $(q=u,d,s,c)$ events.

The events with $D$ meson decays, especially the decay chain $B \rightarrow D^{(*)}\rho^+$ followed by $\rho^+ \to \pi^+\pi^0$ with a high energy photon from the $\pi^0$, give rise to a peak in the signal region of $M_{\rm bc}$.
In order to suppress this background, a $D$ veto is applied for candidates with $M_{X_s} >$\,2.0\,GeV/$c^2$.
$D$ meson candidates of the major 19 hadronic decay modes are reconstructed with combinations of particles used in the $X_s$ reconstruction. 
The event is rejected if any of the $D$ meson candidates falls in a veto window around the $D$ meson mass. 
We set the central value and the width of the veto window depending on the charge of the $D$ candidate and whether or not the $D$ candidate is reconstructed in a mode with a $\pi^0$ or $\eta$ meson: the windows are $1835 < M_{D^0} < 1895$~MeV/$c^2$ and $1840 < M_{D^+} < 1900$~MeV/$c^2$ for the modes without $\pi^0$ or $\eta$, and $1800 < M_{D^0} < 1905$~MeV/$c^2$ and $1805 < M_{D^+} < 1910$~MeV/$c^2$ for the modes with $\pi^0$ or $\eta$.

\begin{figure}[tb]
	\begin{center}
		\includegraphics[width=0.47\textwidth]{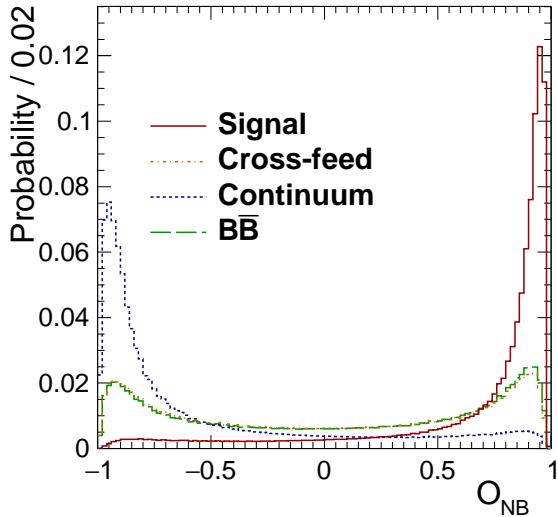}
		\caption{Neural Network output in simulated data that ranges from $-1$ for the $q\bar{q}$ background-like events to $+1$ for the signal-like events. The solid (brown) curve shows signal, the dotted-dashed (orange) curve represents cross-feed, the dashed (blue) curve is $q\bar{q}$ background, and the long dashed (green) curve shows $B\bar{B}$ background.}
	    \label{fig:nboutput}
	\end{center}	
\end{figure}

The continuum background is suppressed using a multivariate analysis with an artificial neural network~\cite{NB},
mostly relying on the difference in topology of continuum (jet-like) and $B\bar{B}$ (spherical) events.
We use the following variables calculated in the CM frame as input parameters to the neural network: 
(1) the cosine of the angle between the $B$ meson candidate momentum and the $z$ axis, 
(2) the likelihood ratio of modified Fox-Wolfram moments~\cite{SFW,KSFW}, 
(3) the cosine of the angle between the thrust axes of the daughter particles of the $B$ candidate and all other particles in the rest of the event~(ROE), 
(4) the thrust value of particles in the ROE,
(5) the sphericity and aplanarity~\cite{sph} of particles in the ROE, 
(6) the cosine of the angle between the first sphericity axes of the $B$ candidate and the particles in the ROE, 
(7) the cosine of the angle between the second sphericity axes of the $B$ candidate and the particles in the ROE, 
(8) the cosine of the angle between the third sphericity axes of the $B$ candidate and the particles in the ROE, 
(9) the cosine of the angle between the first sphericity axis of the particles in the event and the $z$ axis, 
and
(10) the signal probability density for the $\Delta E$ value.
The neural network is trained with signal and {$q\bar{q}$}-background MC simulation events with  2.2\,GeV/$c^2<M_{X_s}<2.8$\,GeV/$c^2$.
We obtain a neural network output (${\cal{O}}_{\rm NB}$) between $-1$ and $+1$, which can discriminate the continuum background from signal events. The distribution of ${\cal{O}}_{\rm NB}$ for simulated samples is shown in Fig.~\ref{fig:nboutput}.
The ${\cal{O}}_{\rm NB}$ value is required to be greater than 0.87 in order to maximize the signal significance in the range 2.2 \,GeV/$c^2<M_{X_s}<2.8$\,GeV/$c^2$, in which the continuum background is the most substantial.
This selection suppresses about 98.5\% of the $q\bar{q}$ background while keeping about 51\% of the signal events in the MC simulation study.
\section{VI. Best Candidates Selection}
After the background suppression, the average number of $B$ candidates per event is about 1.5 for the signal MC simulation sample. In those events we select the candidate 
with the largest ${\cal{O}}_{\rm NB}$. 
This selection keeps about 89\% of the correctly reconstructed signal events while reducing the number of cross-feed events by 45\%, based on the MC simulation study. The ratio of the number of correctly reconstructed signal events to the number of cross-feed events plus correctly reconstructed signal events is improved from 0.67 to 0.77.
Since the average number of $B$ candidates per event for $B\bar{B}$ background is larger, the best candidate selection suppresses this background to 14\%.


\section{VII. Signal Yield Extraction}
To extract the signal yield and physics observables, we perform a simultaneous fit with an extended unbinned maximum likelihood method to eight $M_{\rm bc}$ distributions; five for $B^-$, $B^+$, $\bar{B}^0$, $B^0$, and $B_{\rm fns}$ in the on-resonance data,
and three for charged $B$~($B^-$ and $B^+$), flavor-specific neutral $B$~($\bar{B}^0$ and $B^0$), and $B_{\rm fns}$ in the off-resonance data.
Since the off-resonance data only contain continuum backround, this is useful to constrain the continuum background shape.

To take into account the run by run difference of beam energy, the $M_{\rm bc}$ value is shifted with $E_{\rm beam}^{\rm nom} - E_{\rm beam}^{\rm run}$, where $E_{\rm beam}^{\rm nom}$ is the nominal beam energy set to 5.289~GeV and $E_{\rm beam}^{\rm run}$ is the beam energy for a specific run. By this calibration, the endpoint of the $M_{\rm bc}$ distribution for any run is 5.289~GeV.

The likelihood function consists of probability density functions~(PDFs) for signal, cross-feed, peaking and non-peaking background from $B\bar{B}$ events, and $q\bar{q}$ background. 
All signal and background PDFs are considered for the on-resonance data, while only the $q\bar{q}$ background PDF is used to fit to the off-resonance data.
The signal is modeled with a Crystal Ball function~\cite{CB}:
\begin{equation}
  \nonumber
	f_{\rm CB}(x) = \left \{
	\begin{array}{l}
		\exp\left(-\frac{1}{2}\left(\frac{x-m}{\sigma_{\rm CB}} \right)^2 \right) \ \ \ (\frac{x-m}{\sigma_{\rm CB}}\geq-\alpha) \\
		\frac{\left(\frac{n}{\alpha}\right)^n \exp{({-\frac{1}{2}}\alpha^2})} {\left(\frac{n}{\alpha}-\alpha-\frac{x-m}{\sigma_{\rm CB}} \right)^n} \ \ \ (\frac{x-m}{\sigma_{\rm CB}}<-\alpha),
	\end{array}
	\right.
\end{equation}
where $m$ and $\sigma_{\rm CB}$ are the peak position and width, respectively, and the parameters $\alpha$ and $n$ characterize the non-Gaussian tail. 
The peak position is determined with a large-statistics $B \to D \pi$ data sample. The width is obtained from large simulation samples of $B \to X_s \gamma$ decays corrected for the difference between data and MC simulation decays, which is again obtained using the $B \to D \pi$ samples. The $\alpha$ and $n$ are fixed to the values obtained from signal MC simulation samples.

For the cross-feed background, we construct five histogram PDFs originating from $B^-$, $B^+$, $\bar{B}^0$, $B^0$, and $B_{\rm fns}$ with the signal MC simulation sample. 
The fraction of each cross-feed to the corresponding signal is fixed to the MC simulation value. 

A Gaussian function is used to model the peaking background. 
We consider two types of such backgrounds, originating from $\pi^0$ decays and others dominated by $\eta$ decays. All parameters for the $\pi^0$ peaking background are fixed using the events in the sideband defined as ${\cal{ L }}_{\pi^0}>0.5$ and ${\cal{ L }}_{\eta}<0.2$. The parameters for other peaking background are fixed with the $B\bar{B}$ background MC simulation samples.

The non-peaking background from $B\bar{B}$ events is modeled with an ARGUS function~\cite{ARGUS}:
\begin{equation}
	\footnotesize
        \nonumber
	f_{\rm ARG}(x) = x \biggl\{ 1-\left( \frac{x}{E_{\rm beam}^{\rm CM}}\right) ^2 \biggr\}^{1/2}  \exp\Biggl[ c \biggl\{ 1-\left( \frac{x}{E_{\rm beam}^{\rm CM}}\right) ^2\biggr\}\Biggr],
	\label{eq:argus}
\end{equation}
where $E_{\rm beam}^{\rm CM}$ is fixed to 5.289~GeV and other shape parameters are determined from the MC simulation.
The yields for charged $B$ and neutral $B$ are constrained from the on-resonance and scaled off-resonance events in the sideband defined as $M_{\rm bc}<5.27$~GeV, separately.

For the $q\bar{q}$ background PDF, we use a modified ARGUS function:
\begin{equation}
	\footnotesize
        \nonumber
	f_{\rm ARG}^{\rm mod}(x) = x \biggl\{ 1-\left( \frac{x}{E_{\rm beam}^{\rm CM}}\right) ^2 \biggr\}^p  \exp\Biggl[c_{\rm mod} \biggl\{1-\left(\frac{x}{E_{\rm beam}^{\rm CM}}\right)^2\biggr\}\Biggr],
\end{equation}
where a power parameter, $p$, is introduced to account for the steep slope at low $M_{\rm bc}$. 
The $c_{\rm mod}$ parameter is common for on- and off-resonance data and floated in the fit.
The $p$ parameter is fixed from the $q\bar{q}$ MC simulation samples calibrated with the off-resonance data. 

There are in total 16 free parameters in the simultaneous fit: five signal yields for the five $B$ categories, 
eight $q\bar{q}$ yields (five for on-resonance and three for off-resonance), and three $c_{\rm mod}$ shape parameters for $q\bar{q}$.
Finally, the physics parameters to be extracted can be written in terms of the efficiency corrected signal yields~($N_i$) as:
\begin{eqnarray}
\nonumber
&&   \Delta_{0-} = \frac{\frac{\tau_{B^+}}{\tau_{B^0}}\frac{f_{+-}}{f_{00}} (N_{\bar{B}^0} + N_{B^0} + N_{B_{\rm fns}} ) - (N_{B^-} + N_{B^+})}
                     {\frac{\tau_{B^+}}{\tau_{B^0}}\frac{f_{+-}}{f_{00}} (N_{\bar{B}^0} + N_{B^0} + N_{B_{\rm fns}} ) + (N_{B^-} + N_{B^+})},\\
\nonumber
&&   A_{CP}^{\rm C} = \frac{N_{B^-} - N_{B^+}}{N_{B^-} + N_{B^+}},\\
\nonumber
&&   A_{CP}^{\rm N} = \frac{N_{\bar{B}^0} - N_{B^0}}{N_{\bar{B}^0} + N_{B^0}},\\ 
&&   A_{CP}^{\rm tot} = \frac{(N_{B^-} + N_{\bar{B}^0}) - (N_{B^+} + N_{B^0})}{(N_{B^-} + N_{\bar{B}^0}) + (N_{B^+} + N_{B^0})},
\end{eqnarray}
where 
$A_{CP}^{\rm N}$, $A_{CP}^{\rm C}$ and $A_{CP}^{\rm tot}$ are direct $CP$ violation parameters for neutral, charged and combined $B \to X_s \gamma$ decays,
$\tau_{B^+}/\tau_{B^0}$ is the lifetime ratio fixed to the PDG value~\cite{PDG2018}, and $f_{+-}/f_{00}$ is the production ratio of $B^+B^-$ to $B^0\bar{B}^0$ from  $\Upsilon(4S)$ decays also fixed to the PDG value~\cite{PDG2018}~\cite{Jung:2015yma}.
The fitting procedure is validated using the full MC simulation samples and an ensemble test based on toy MC simulation samples.

\section{VIII. Calibration of $X_s$ Fragmentation Model}
Since the signal efficiency depends on specific decay modes, the fragmentation model in the inclusive MC simulation is calibrated to that of the data to reduce the systematic uncertainties associated to modeling. The final states are divided into ten categories, defined in Table~\ref{tab:modecategory}, and four $M_{X_s}$ regions are defined ([1.15,1.5]GeV/$c^2$, [1.5,2.0]GeV/$c^2$, [2.0,2.4]GeV/$c^2$ and [2.4,2.8]GeV/$c^2$) to calibrate the fragmentation model. We adopt the same calibration method as described in Ref.~\cite{Saito:2014das}.

\section{IX. Systematic Uncertainties}
We evalulate the systematic uncertainties associated with particle detection efficiencies, charge asymmetries in particle detection, selections, physics parameters, $X_s$ fragmentation model, background $A_{CP}$ and $\Delta_{0-}$, fixed parameters in the fit, fitter bias, and MC simulation statistics. We list the systematic uncertainties in Table~\ref{tab:syst}.

We estimate the tracking efficiency uncertainty using partially reconstructed $D^{*+} \to D^0 \pi^+$, $D^0 \to K_S^0 \pi^+ \pi^-$ events. 
The uncertainties due to kaon and pion identifications are evaluated with clean kaon and pion samples in $D^{*+} \to D^0 \pi^+$, followed by $D^0 \to K^- \pi^+$. 
We determine the uncertainty due to $\pi^0$ reconstruction by taking the ratio of the efficiencies of $\eta \to 3\pi^0$ to $\eta \to \pi^+ \pi^- \pi^0$ or $\eta \to \gamma \gamma$. 
The uncertainty due to $K_S^0$ reconstruction is evaluated by checking the efficiency of $K_S^0 \to \pi^+ \pi^-$ as functions of flight length, transverse momentum of $K_S^0$, and polar angle of $K_S^0$. 

We measure the charged-pion detection asymmetry using reconstructed $B \to X_s \gamma$ candidates (with charged pion in the final state) in the sideband defined as ${\cal{O}}_{\rm NB}<0$. The charged kaon detection asymmetry is measured using a large clean kaon sample from $D^0 \to K^{-} \pi^{+}$ decay, where the pion detection asymmetry in the decay is subtracted with pions from $D_s^{+} \to \phi \pi^{+}$ decays~\cite{Ko:2012uh}.

The uncertainty due to possible mismodeling of the $\Delta E$ distribution we estimate by inflating the $\Delta E$ width and shifting the mean value. 

We evalulate the uncertainties due to $f_{+-}/f_{00}$ and lifetime ratio by changing these values by $\pm1\sigma$ from the nominal PDG values~\cite{PDG2018}.

The uncertainty due to the fragmentation model we determine by varying the decay channel proportions by their respective uncertainties.
The exceptions are the proportions for $K4\pi$ and $K2\pi^0$ in 2.0\,GeV/$c^2 <M_{X_s}< 2.4$\,GeV/$c^2$ and all the modes in 2.4\,GeV/$c^2 <M_{X_s}< 2.8$\,GeV/$c^2$, where we use the proportions in MC simulation and a variation of $\pm50\%$ as uncertainty.
The fragmentation uncertainties for each $M_{X_s}$ bin are obtained by summing in quadrature the changes for each of the ten mode categories.

Since the threshold between $K^*$ and the inclusive $X_s$ used in the MC simulation is fixed at 1.15\,GeV/$c^2$, we change this boundary to 1.10\,GeV/$c^2$ and 1.20\,GeV/$c^2$ to evaluate the uncertainty due to the threshold.

The proportion of missing final states that are not included in our reconstructed modes affects the reconstruction efficiency. 
We evaludate the uncertainty on the relative proportion of each of the 38 measured final states by varying the parameters of the fragmentation model used in the calibration of the MC simulation within their allowed ranges as determined from data.
We take the difference from the nominal value as the systematic uncertainty on the missing fraction.

We evaludate the uncertainties due to $A_{CP}$ and $\Delta_{0-}$ in the background decays by changing the $A_{CP}$ and $\Delta_{0-}$ values by $\pm1\sigma$ from the nominal PDG values~\cite{PDG2018}; if neither $A_{CP}$ nor $\Delta_{0-}$ are measured, we assign $\pm100\%$ uncertainties.

We evaluate the uncertainties due to tail parameters, $\alpha$ and $n$, in the signal PDF by floating in turn each of the fixed shape parameters in the fit while fixing the other shape parameters to their nominal values. Then the two uncertainties are added in quadrature. Since the $\alpha$ and $n$ are anti-correlated, this procedure conservatively estimates the uncertainties.
The uncertainties due to the other fixed parameters in the signal PDF are evaluated by varying them by $\pm1\sigma$ from the nominal values.
The uncertainty due to cross-feed is caused by two sources; one is multiplicity of hadrons in the other $B$ meson decays, the other is fragmentation model for signal. Both changes the shape and yield of cross-feed. The former is evaluated with MC simulation by changing the multiplicities of $\pi^{\pm}$, $\pi^0$, $K^{\pm}$, $K^0$ and $\eta$ in the other $B$ meson decays MC simulation from the nominal values to PDG values taking into account their uncertainties~\cite{PDG2018}. The latter is determined with MC simulation by varying the decay channel proportions by their respective uncertainties.
We estimate the uncertainty due to the $p$ parameter in $q\bar{q}$ background PDF by changing the parameter by $\pm1\sigma$ as obtained from the fit to the off-resonance data.
To evaluate the uncertainty due to the peaking background from $\pi^0$ decays, we vary the parameter values by $\pm1\sigma$ as determined from the sideband data.
The systematic uncertainties of other peaking backgrounds, which are subleading to the $\pi^0$ backgrounds, we evaluate by changing the normalizations 
by $\pm20\%$ which is about twice larger than the uncertainties of the corresponding branching fractions.

We check for possible bias in the fit by performing a large number of pseudo-experiments. In the study, we observe small biases which we add to the systematic uncertainty.

We also take into account the statistical uncertainty of the efficiency estimated with MC simulation samples as systematic uncertainty.

The systematic uncertainties due to efficiencies and background $\Delta_{0-}$ are only relevant for $\Delta_{0-}$ and $A_{CP}^{\rm tot}$ since these cancel out by taking the $CP$ asymmetry in the other observables. The systematic uncertainties due to physics parameters to convert the signal yields to decay widths are only relevant for $\Delta_{0-}$. The systematic uncertainties due to charged particle detection asymmetry and background $A_{CP}$ are only relevant for $A_{CP}$ as they cancel out for the $CP$-averaged observable $\Delta_{0-}$.
The largest and dominant systematic uncertainty for $\Delta_{0-}$ is due to $f_{+-}/f_{00}$. 
The dominant systematic sources for $\Delta A_{CP}$ and $A_{CP}$ are due to peaking background from $\pi^0$ decays and charge asymmetries in particle detection.

\begin{table}[htb]
\centering
\caption{Absolute systematic uncertainties for $\Delta_{0-}$, $\Delta A_{CP}$ and $A_{CP}$ in percent.}
\label{tab:syst}
\begin{tabular}{l|cccccc}
\hline \hline
\newlength{\myheight}
\setlength{\myheight}{4mm}
\rule{0cm}{\myheight}Source     & $\Delta_{0-}$ & $\Delta A_{CP}$ & $A_{CP}^{\rm C}$ & $A_{CP}^{\rm N}$ & $A_{CP}^{\rm tot}$ & $\bar{A}_{CP}$  \\
\hline
tracking                        & $\pm0.02 $ &  --   &   --   &   --   &  $<0.01$  &   --  \\
$K/\pi$ ID                      & $\pm 0.05   $ &   --   &   --   &   --   &   $<0.01$   &   --  \\
$\pi^0/\eta$ recon.             & $\pm  0.01   $ &  --  &   --  &   --  &   $<0.01$   &   --  \\
$K_S^0$ recon.                  & $\pm 0.01   $ &   --  &   --  &  --  &   $<0.01$   &   --  \\
detection asym.             &  -- & $\pm  0.39  $ & $\pm  0.11  $ & $\pm  0.29  $ & $\pm  0.05  $ & $\pm  0.10  $\\
$\Delta E$ selection           & ${}^{+0.03}_{-0.06}$   &  --   &   --   &   --   &   $<0.01$   &   --  \\
$f_{+-}/f_{00}$                  & $\pm  1.15 $  &   --   &   --   &   --   &   --   &   --  \\
lifetime ratio                 & $\pm  0.19 $  &   --   &   --   &   --   & --   &   --  \\
fragmentation                  & $\pm  0.58 $ & -- & -- & -- & $\pm 0.01$   & --  \\
$K^{\ast}$-$X_s$ transition           & $\pm  0.13 $  &   --   &   --   &   --   &  $<0.01$  &   --  \\
missing fraction                   & $ \pm 0.02 $  & -- & -- & -- & $<0.01$   & --  \\
background $A_{CP}$             & -- & $\pm 0.04  $ & $\pm  0.03  $ & $\pm  0.04  $ & $\pm  0.02  $ & $\pm  0.02  $\\
background $\Delta_{0-}$        & $\pm  0.01  $ & -- & -- & -- & $<0.01$   & --  \\
fixed parameters               & ${}^{+0.74}_{-0.65}$ & ${}^{+0.64}_{-0.61}$ & ${}^{+0.30}_{-0.28}$ & $ {}^{+0.34}_{-0.36}$ & $ \pm 0.07 $ & ${}^{+0.07}_{-0.06}$\\
fitter bias                  & $ {}^{+0.08}_{-0.07} $ & $ {}^{+0.11}_{-0.07} $ & $ {}^{+0.04}_{-0.00} $ & $ {}^{+0.10}_{-0.09} $ & $ {}^{+0.05}_{-0.02} $ & $ {}^{+0.06}_{-0.03} $\\ 
MC sim. stat.                     & $\pm  0.03  $ & --    & --  & -- & $<0.01$ & -- \\ 
\hline
total                        & ${}^{+1.51}_{-1.47}$ & ${}^{+0.76}_{-0.73}$ & ${}^{+0.32}_{-0.30}$ & ${}^{+0.46}_{-0.47}$ & ${}^{+0.11}_{-0.09}$ & ${}^{+0.13}_{-0.12}$ \\ 
\hline \hline
\end{tabular}
\end{table}

\section{X. Results}
\begin{figure*}[htbp]
    \begin{center}
    \includegraphics[width=0.4\textwidth]{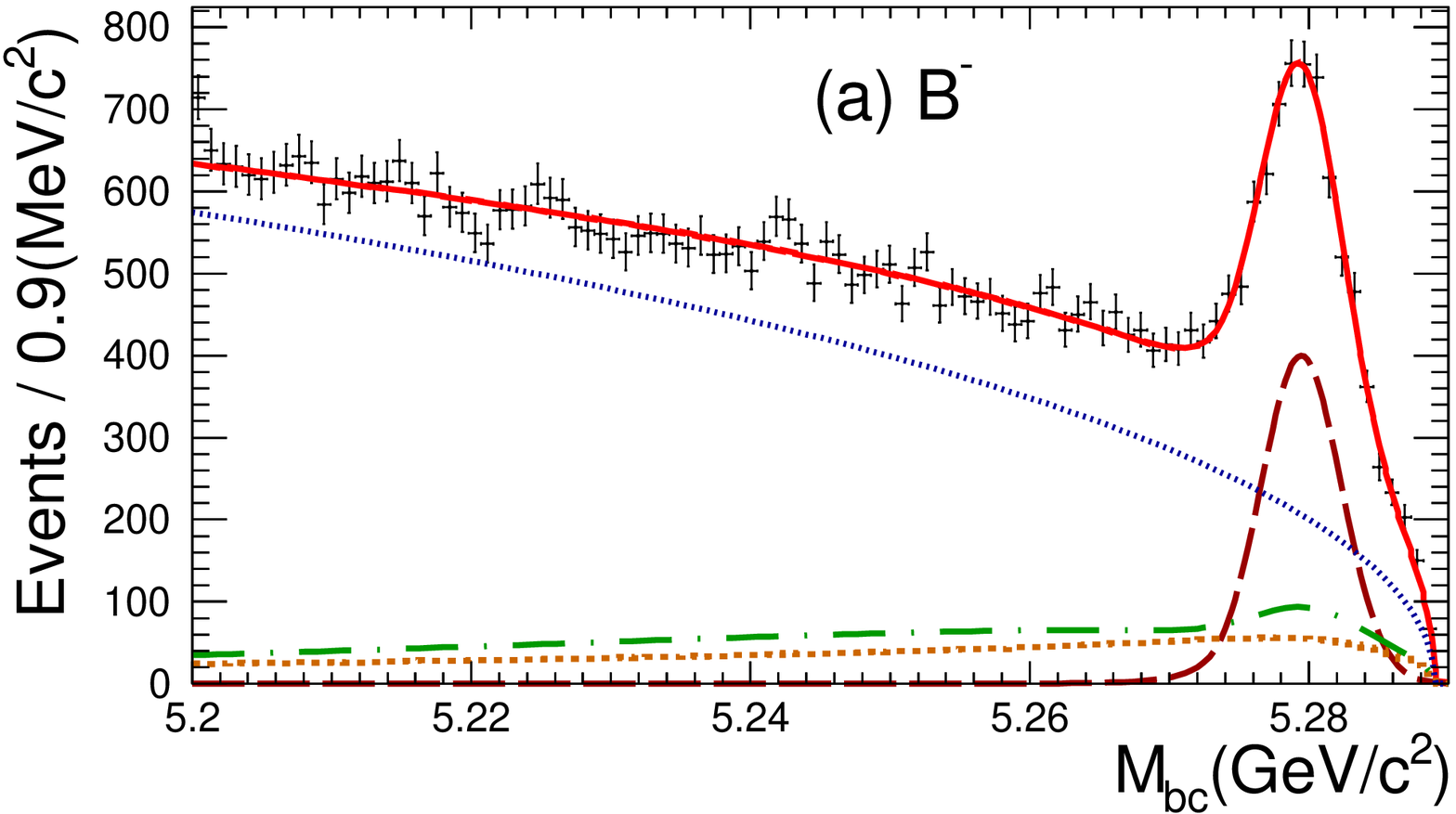}
    \includegraphics[width=0.4\textwidth]{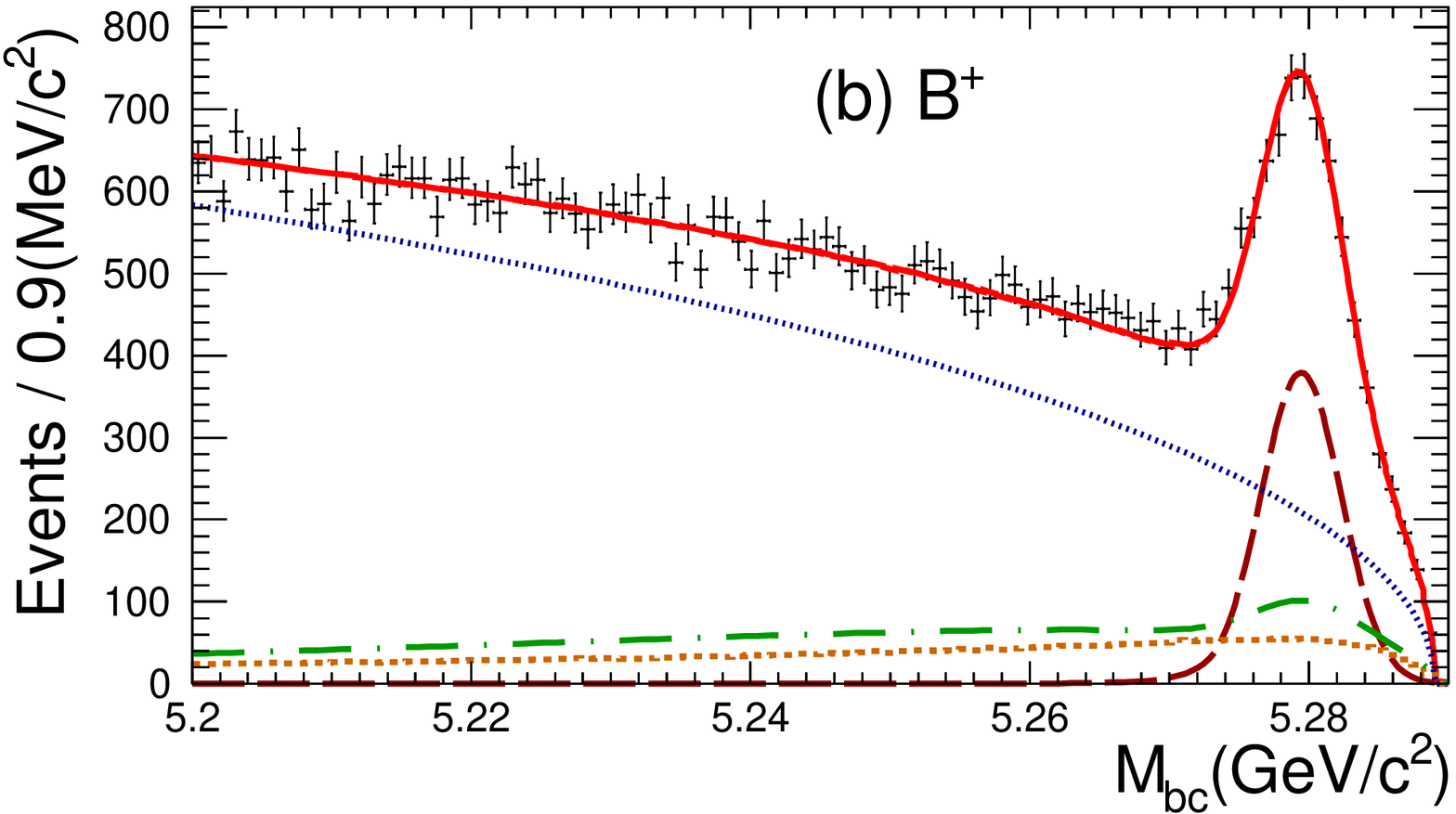}\\
    \includegraphics[width=0.4\textwidth]{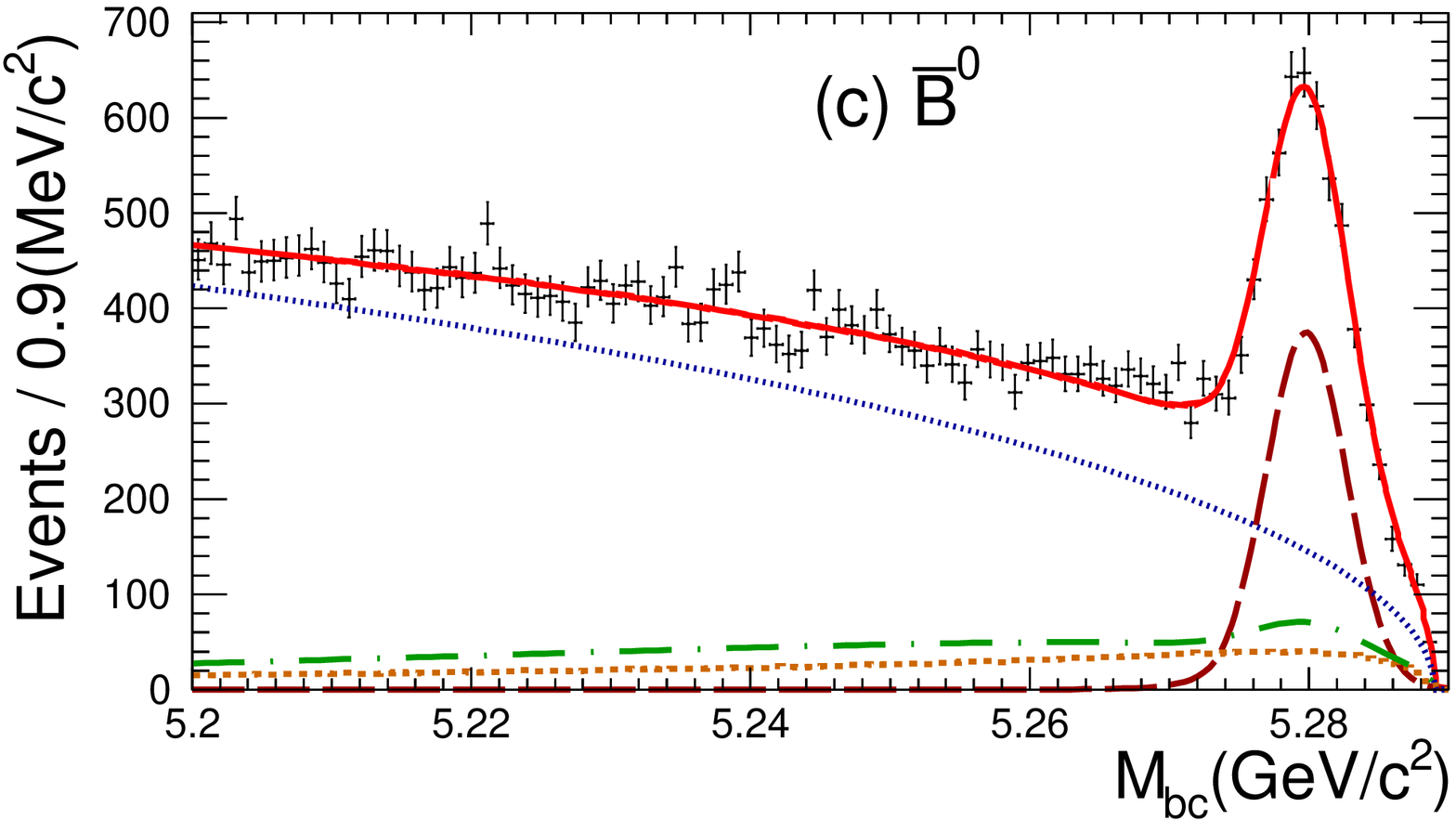}
    \includegraphics[width=0.4\textwidth]{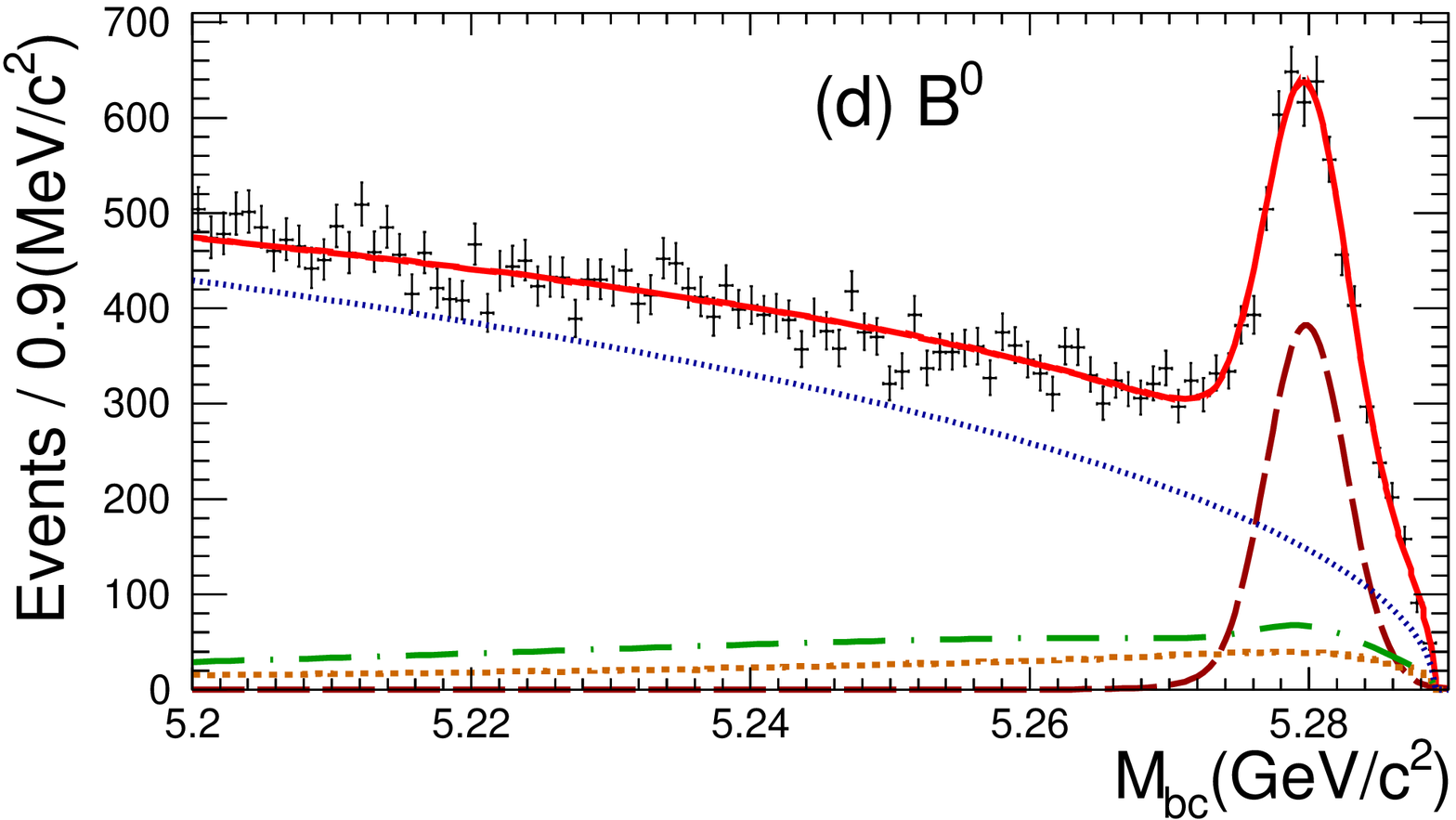}\\
    \includegraphics[width=0.4\textwidth]{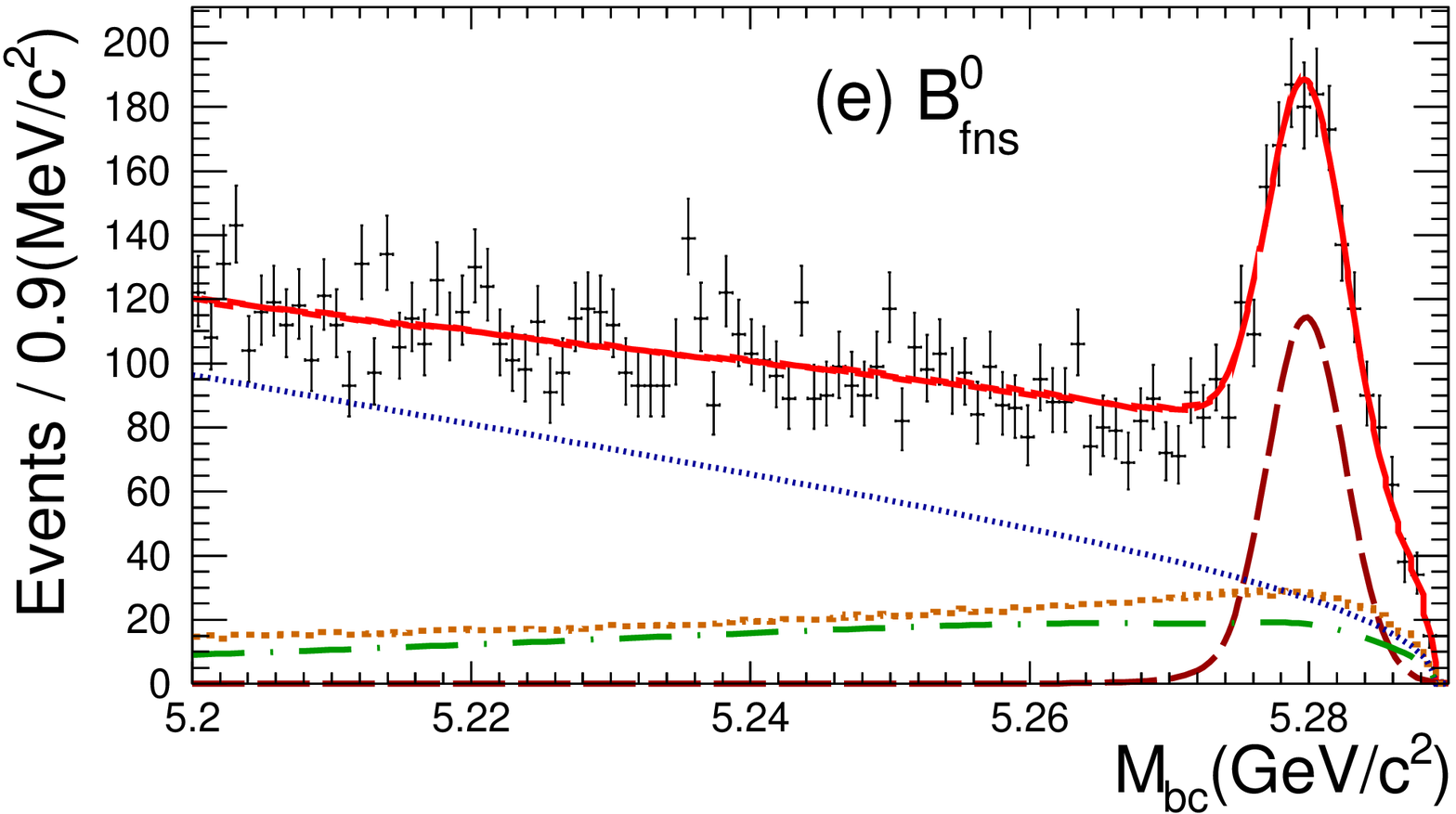}\\
    \includegraphics[width=0.4\textwidth]{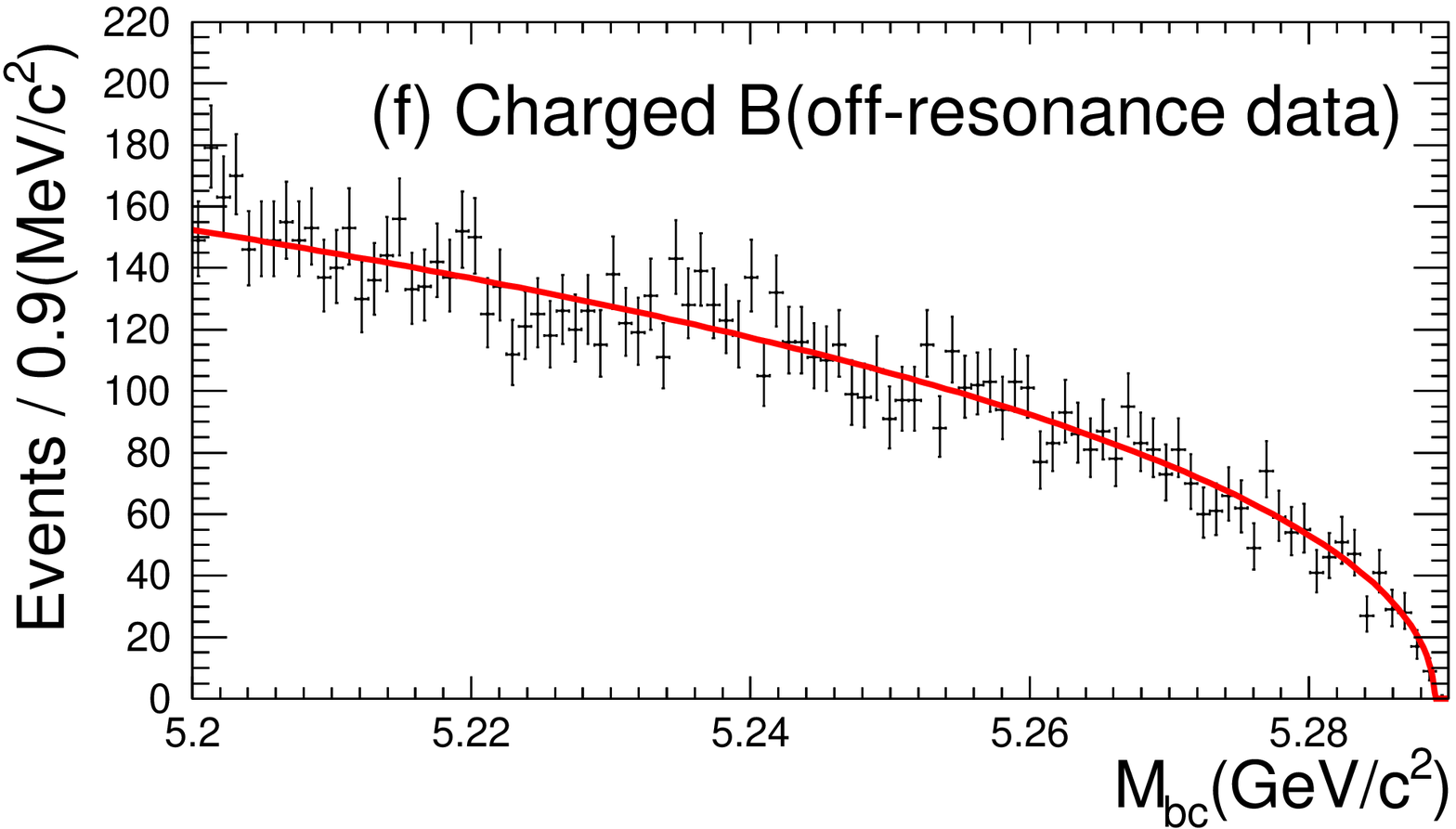}
    \includegraphics[width=0.4\textwidth]{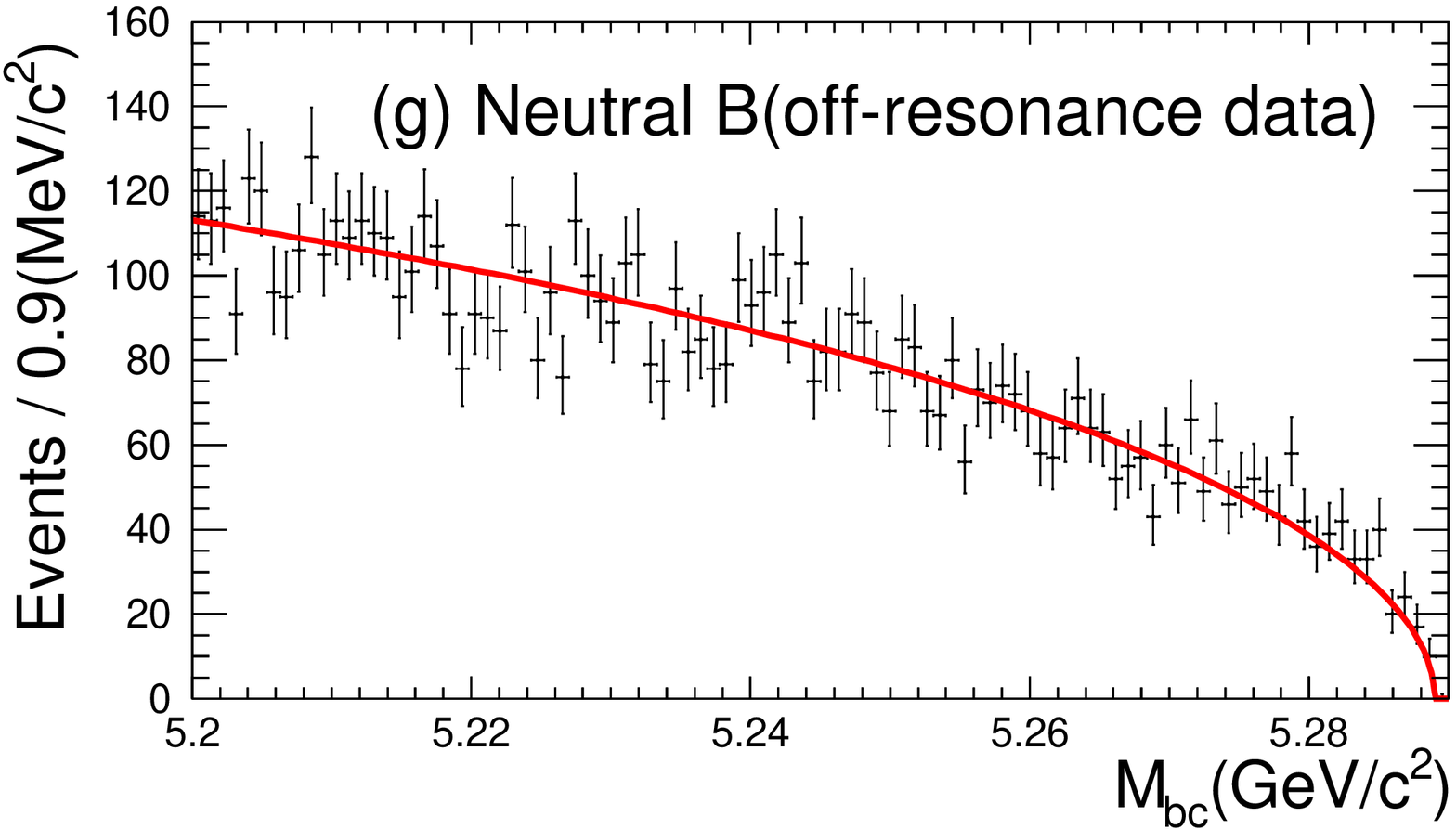}\\
    \includegraphics[width=0.4\textwidth]{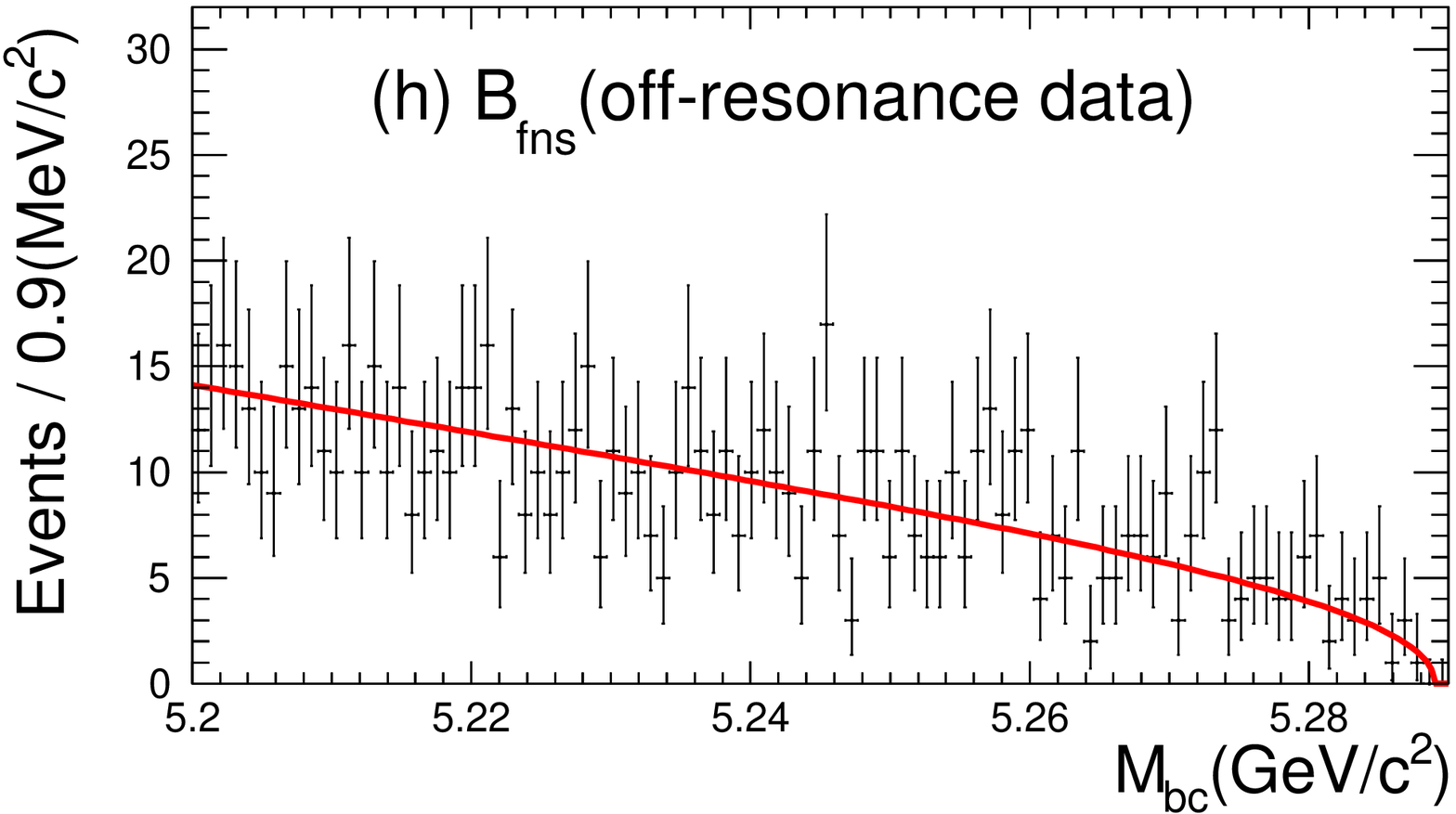}
    \end{center}
    \caption { $M_{\rm bc}$ distributions for (a) $B^-$, (b) $B^+$, (c) $\bar{B}^0$, (d) $B^0$, and (e) $B_{\rm fns}$ in on-resonance data, and (f) charged $B$, (g) neutral $B$ and (h) $B_{\rm fns}$ in off-resonance data. The points with error bars show the data and the lines show different contributions as obtained from the fit. The long dashed (brown) curves represent signal, the dotted (blue) curves show continuum, the dotted-dashed (green) curves are $B\bar{B}$ background, the dashed (orange) curves show cross-feeed, and solid (red) curves are the sum of all contributions.
             }
    \label{fig:mbcfit}
\end{figure*}
We perform a simultaneous fit to eight $M_{\rm bc}$ distributions shown in Figure~\ref{fig:mbcfit}, with the PDFs as described above, to extract the following results
\begin{eqnarray}
\nonumber
  \Delta_{0-}   &=& (-0.48 \pm 1.49  \pm 0.97  \pm 1.15)\%,\\
\nonumber
  \Delta A_{CP} &=& (+3.69 \pm 2.65  \pm 0.76)\%,\\
\nonumber
  A_{CP}^{\rm C}    &=& (+2.75 \pm 1.84  \pm 0.32)\%, \\
\nonumber
  A_{CP}^{\rm N}    &=& (-0.94 \pm 1.74  \pm 0.47)\%,\\
\nonumber
  A_{CP}^{\rm tot}  &=& (+1.44 \pm 1.28  \pm 0.11)\%,\\
\nonumber
  \bar{A}_{CP} &=& (+0.91 \pm 1.21  \pm 0.13)\%,
\end{eqnarray}
where the first uncertainty is statistical, the second is systematic, and the third for $\Delta_{0-}$ is due to $f_{+-}/f_{00}$. The fit results for the signal yields are summarized in Table~\ref{tab:result}. The correlation matrix of the six observables is given in Table~\ref{tab:cor}.
The $\chi^2$ and number of degrees of freedom in the simultaneous fit calculated from the data points and fit curves in Fig.~\ref{fig:mbcfit} are 728 and 784, respectively.
The measured $\Delta_{0-}$ is consistent with zero and the precision is better than that of BaBar by a factor of three~\cite{Aubert:2005cua}. Thus, this measurement can be used to constrain the resolved photon contribution in $B \to X_s \gamma$ as
\begin{eqnarray}
\nonumber
\frac{{\cal{B}}_{\rm RP}^{78}}{\cal{B}} &\simeq& (+0.16\pm0.50\pm0.32\pm0.38\pm0.05)\%,
\label{eqn:resultrp}
\end{eqnarray}
where the last uncertainty is associated with $SU(3)$ flavor-symmetry breaking.
This result improves the prediction of the branching fraction. 
The result for $\Delta A_{CP}$ is consistent with zero, as predicted in the SM, thus the measurement can be used to constrain NP models, for example, in the supersymmetry model described in Ref.~\cite{Endo:2017ums}, this excludes the parameter space for squark mass below 5.0~TeV$/c^2$.

We checked $M_{X_s}$ dependences of the observables and find no dependences except $\Delta_{0-}$ in $K^*$ mass region is larger than the measurement and is consistent with the world average~\cite{PDG2018}.
 
\begin{table}[htb]
\centering
\caption{Signal yields~($N_S$) and efficiencies ($\epsilon$). The uncertainties for $N_S$ are statistical. The uncertainties for $\epsilon$ include systematic uncertainties.}
\label{tab:result}
\begin{tabular}{l|cc}
\hline \hline
Mode & $N_S$                        & $\epsilon$ [\%] \\
\hline
$ B^-       $               & $ 3243 \pm 85 $    & $ 2.21   \pm 0.12 $  \\
$ B^+       $               & $ 3074 \pm 86 $    & $ 2.23   \pm 0.12 $ \\
$ \bar{B}^0 $               & $ 3038 \pm 78 $    & $ 2.42   \pm 0.14 $  \\
$ B^0       $               & $ 3102 \pm 79 $    & $ 2.46   \pm 0.14 $  \\
$ B_{\rm fns} $               & $  902 \pm 42 $    & $ 0.375  \pm 0.023 $  \\
\hline \hline
\end{tabular}
\end{table}

\begin{table}[htb]
\centering
\caption{The correlation matrix for the six observables.}
\label{tab:cor}
\begin{tabular}{l|rrrrrr}
\hline \hline
\setlength{\myheight}{4mm}
\rule{0cm}{\myheight}
                   & $\Delta_{0-}$ & $\Delta A_{CP}$ & $A_{CP}^{\rm C}$ & $A_{CP}^{\rm N}$ & $A_{CP}^{\rm tot}$ & $\bar{A}_{CP}$  \\
\hline
$\Delta_{0-}$       & $ 1.00$  & $ 0.07 $ & $  0.06 $ & $  -0.05 $ & $ -0.01 $ & $  0.01 $ \\
$\Delta A_{CP}$     & $ 0.07$  & $ 1.00 $ & $  0.70 $ & $ -0.68 $ & $ 0.29 $ & $  0.03  $ \\
$A_{CP}^{\rm C}$      & $ 0.06$  & $ 0.70 $ & $  1.00 $ & $ -0.12 $ & $ 0.91 $ & $  0.74  $ \\
$A_{CP}^{\rm N}$      & $-0.05$  & $-0.68 $ & $ -0.12 $ & $  1.00 $ & $ 0.47 $ & $  0.72  $ \\
$A_{CP}^{\rm tot}$     & $-0.01$ & $ 0.29 $ & $  0.91 $ & $  0.47 $ & $ 1.00 $ & $  0.94 $  \\
$\bar{A}_{CP}$      & $ 0.01$ & $ 0.03 $ & $  0.74 $ & $  0.72 $ & $ 0.94 $ & $  1.00 $  \\
\hline \hline
\end{tabular}
\end{table}

\section{XI. Confidence level Intervals}
\begin{figure}[htbp]
    \begin{center}
    \includegraphics[width=0.47\textwidth]{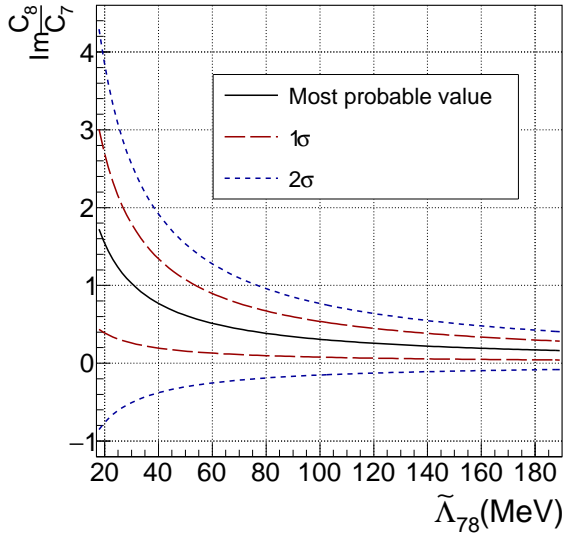}
    \caption {The solid (black), dashed (red) and dotted (blue) curves show the most probable value, $1\sigma$ and $2\sigma$ confidence intervals for ${\rm Im}(C_8/C_7)$ as a function of $\tilde \Lambda_{78}$, respectively. The range of $\tilde \Lambda_{78}$ is chosen to be 17MeV $< \tilde \Lambda_{78} <$ 190MeV.}
    \label{fig:contour}
    \end{center}
\end{figure}

 From our measurement of $\Delta A_{CP}$, we set confidence intervals on ${\rm Im}(C_8/C_7)$ based on Eq.~(\ref{eqn:dacp}).
The hadronic parameter $\tilde \Lambda_{78}$ has a large uncertainty and the range is estimated as 17~MeV $< \tilde \Lambda_{78} <$ 190~MeV with a vacuum insertion approximation~\cite{Benzke:2010tq}. 
Since an uncertainty of $\tilde \Lambda_{78}$ is hard to estimate, we set the 1$\sigma$ and 2$\sigma$ confidence level invervals for ${\rm Im}(C_8/C_7)$ as a function of $\tilde \Lambda_{78}$ within the range described above as shown in Figure~\ref{fig:contour}. Our result constrains ${\rm Im}(C_8/C_7)$ in the positive region better than the only previously available measurement from BaBar~\cite{Lees:2014uoa}, and gives a strong constraint on NP models~\cite{Endo:2017ums}. If we take the average value of $\tilde \Lambda_{78}=89$~MeV as a benchmark~\cite{Malm:2015oda}, the 2$\sigma$ confidence intervals is $-0.17 < {\rm Im}(C_8/C_7)<0.86$.

\section{XII. Conclusion}

In summary, we have measured the isospin asymmetry and the difference of the direct $CP$ asymmetries between charged and neutral $B \to X_s \gamma$ decays with a sum-of-exclusive technique based on a sample of ${772 \times 10^6}$ $B\bar{B}$ pairs with an assumption that the observables have no dependence on the specific decay modes nor on $M_{X_s}$. The measurement of $\Delta_{0-}$ is consistent with zero and can constrain the resolved photon contribution in $B \to X_s \gamma$, which will improve the prediction of the branching fraction. 
The result of $\Delta A_{CP}$ is consistent with zero as predicted in the SM, enabling constraints on NP models.
Our measurements of the $CP$ asymmetries are consistent with zero, and also with the SM predictions. All the results are the most precise to date and will be useful for constraining the parameter space in NP models. 
Current $A_{CP}$ and $\Delta A_{CP}$ measurements are dominated by
the statistical uncertainty; thus, the upcoming Belle~II
experiment will further reduce the uncertainty. To improve the isospin asymmetry
at Belle~II, reduction of the dominant uncertainty due to $f_{+-}/f_{00}$ is essential, and can be performed at both Belle and Belle~II.

\section{ACKNOWLEDGMENTS}

Authors would like to thank M.~Misiak, G.~Paz and M.~Endo for fruitful discussions.
A.~I. is supported by the Japan Society for the Promotion of Science (JSPS) Grant No.~16H03968.
We thank the KEKB group for the excellent operation of the
accelerator; the KEK cryogenics group for the efficient
operation of the solenoid; and the KEK computer group,
the National Institute of Informatics, and the 
Pacific Northwest National Laboratory (PNNL) Environmental Molecular Sciences Laboratory (EMSL) computing group for valuable computing
and Science Information NETwork 5 (SINET5) network support.  We acknowledge support from
the Ministry of Education, Culture, Sports, Science, and
Technology (MEXT) of Japan, the Japan Society for the 
Promotion of Science (JSPS), and the Tau-Lepton Physics 
Research Center of Nagoya University; 
the Australian Research Council;
Austrian Science Fund under Grant No.~P 26794-N20;
the National Natural Science Foundation of China under Contracts
No.~11435013,  
No.~11475187,  
No.~11521505,  
No.~11575017,  
No.~11675166,  
No.~11705209;  
Key Research Program of Frontier Sciences, Chinese Academy of Sciences (CAS), Grant No.~QYZDJ-SSW-SLH011; 
the  CAS Center for Excellence in Particle Physics (CCEPP); 
Fudan University Grant No.~JIH5913023, No.~IDH5913011/003, 
No.~JIH5913024, No.~IDH5913011/002;                        
the Ministry of Education, Youth and Sports of the Czech
Republic under Contract No.~LTT17020;
the Carl Zeiss Foundation, the Deutsche Forschungsgemeinschaft, the
Excellence Cluster Universe, and the VolkswagenStiftung;
the Department of Science and Technology of India; 
the Istituto Nazionale di Fisica Nucleare of Italy; 
National Research Foundation (NRF) of Korea Grants No.~2014R1A2A2A01005286, No.2015R1A2A2A01003280,
No.~2015H1A2A1033649, No.~2016R1D1A1B01010135, No.~2016K1A3A7A09005 603, No.~2016R1D1A1B02012900; Radiation Science Research Institute, Foreign Large-size Research Facility Application Supporting project and the Global Science Experimental Data Hub Center of the Korea Institute of Science and Technology Information;
the Polish Ministry of Science and Higher Education and 
the National Science Center;
the Ministry of Education and Science of the Russian Federation and
the Russian Foundation for Basic Research;
the Slovenian Research Agency;
Ikerbasque, Basque Foundation for Science, Basque Government (No.~IT956-16) and
Ministry of Economy and Competitiveness (MINECO) (Juan de la Cierva), Spain;
the Swiss National Science Foundation; 
the Ministry of Education and the Ministry of Science and Technology of Taiwan;
and the United States Department of Energy and the National Science Foundation.

%


\begin{thebibliography}{99}
\bibitem{CC}
Throughout this paper, the inclusion of the charge conjugate mode decay is implied unless otherwise stated.
\end{thebibliography}
\end{document}